# Preparation and Characterization of $Ni_xMn_{0.25-x}Mg_{0.75}Fe_2O_4$ Nano-ferrite as $NO_2$ Gas Sensing Material

**Hussein I. Mahdi [1], Nabeel A. Bakr [2], Tagreed M. Al-Saadi [3]**

[1,2] Department of Physics, College of Science, University of Diyala, Diyala, IRAQ

[3] College of Education for Pure Science, Ibn Al Haitham, University of Bagdad, Bagdad, IRAQ

[*]Corresponding author: sciphydr2110@uodiyala.edu.iq

## Abstract

$Ni_xMn_{0.25-x}Mg_{0.75}Fe_2O_4$ nano-ferrites (where x = 0.00, 0.05, 0.10, 0.15 and 0.20) were produced via sol-gel auto-combustion technique. Investigations were done into how the incorporation of Ni ions affects the $Mn_{0.25}Mg_{0.75}Fe_2O_4$ ferrite's structure, morphological, magnetic, and $NO_2$ gas sensing features. All the samples are single-phase, based on the structural study utilizing the X-ray diffraction (XRD) pattern. In terms of the structure of the cubic spinel, according to the XRD study, the crystallite sizes range from 24.30 to 28.32 nm, indicating nano-crystallinity. The synthesis of spherical nanoparticles with a small modification in particle size distribution was verified via FE-SEM images. The study found that the size of particles is tiny enough to act superparamagnetically. The area of hysteresis loop is almost non-existing, thus reflecting typical soft magnetic materials according to magnetic measurements by VSM carried out at room temperature. Furthermore, the conductance responses of the $Ni_xMn_{0.25-x}Mg_{0.75}Fe_2O_4$ nano-ferrite were measured by exposing the ferrite to oxidizing ($NO_2$) gas at different operating temperatures. The results show that the sensor boasts shorter response and recovery times, as well as a higher sensitivity 707.22% of the sample (x=0.20) for nano-ferrite.

**Keyword:** Mn-Mg ferrite, Ni ions substitution, sol- gel auto-combustion technique, XRD, VSM, $NO_2$ gas sensor.

## 1. Introduction

Because chemical sensors may control emissions and identify dangerous contaminants, their demand has risen dramatically. The most promising chemical sensors are metal oxide semiconductor ones since they offer several benefits like low cost, compact size, low power consumption, and online operation. They have received extensive research for a long time because they are very suitable with microelectronic processes [1]. Utilization of nanocrystalline materials for gas sensing have recently sparked a great deal of curiosity [2]. Ferrites have proven to be effective materials for gas semiconductor detectors [3]. Whenever a semiconductor gas sensor is exposed to various gas environments, it acquires the ability to modify the conductivity of the detecting material.

The surface-controlled technique of gas sensing depends on the interaction among both gas molecules to be identified and adsorbed oxygen. The operating temperature, the type of gas being used, and the type of detector all affect how the detector responds to gas [4]. The oxides having a structural formula of $AB_2O_4$ are significant for gas detection purposes and were studied for the identification of both oxidizing and reducing gases. These oxides are preferred above all spinel-type metal oxide semiconductor detector, due to the magnetic materials used in high frequency applications as micro-electronic/magnetic devices [5]. The most exciting features of spinel ferrites for gas detecting are their chemical makeup and structure, in which transition or post-transition cations occupy two different cation positions [6]. The spinel ferrites, including $MgFe_2O_4$, $ZnFe_2O_4$, $MnFe_2O_4$, $NiFe_2O_4$, and $CoFe_2O_4$, have shown excellent sensitivity for a wide range of gases due to their stability in thermal and chemical atmospheres, quick reaction and recovery times, inexpensive, and straightforward electronic structures [7,8]. Magnesium ferrite is specifically among the most significant ferrites due to its low magnetic and dielectric losses, high resistivity, and other properties that make it an essential component in catalytic reactions, detectors, and adsorption [9]. Depending on the preferred energies for divalent and trivalent ions in the spinel structure, it possesses an inverse spinel structure with $Mg^{2+}$ ions in octahedral sites and $Fe^{3+}$ ions equally divided over tetrahedral and octahedral sites [10].

The sol-gel, molten-salt approach, hydrothermal, co-precipitation, and microemulsion techniques were all employed to obtain nano-sized spinel ferrite powder [11,12]. Among the numerous techniques, the sol-gel technique is a convenient, environmentally friendly, and low-cost technique for synthesizing ferrites at relatively low temperatures in a short period of time [13].

Doping is a significant and successful method for fine-tuning the required properties of semiconductors [14,15]. The dopant might improve the gas-sensing characteristics of metal-oxide semiconductors by modifying the energy-band structure, improving the morphology and surface-to-volume ratio, and developing extra active centers at the grain boundaries [16].

In the present work, we report the synthesis of $Ni_xMn_{0.25-x}Mg_{0.75}Fe_2O_4$ nano-ferrite by using a simple sol-gel auto-combustion technique and its application as $NO_2$ gas sensor has been systematically investigated, where the results are presented and discussed.

## 2. Experimental Part

### 2.1. Materials and method

The general formula of the spinel ferrite of $Ni_xMn_{0.25-x}Mg_{0.75}Fe_2O_4$ (where x = 0.00, 0.05, 0.10, 0.15 and 0.20) has been produced via sol-gel auto-combustion technique. Analytical-grade materials of ferric nitrate nonahydrate $Fe(NO_3)_3.9H_2O$, magnesium nitrate hexahydrate $Mg(NO_3)_2.6H_2O$, manganese nitrate monohydrate $Mn(NO_3)_2.H_2O$, and nickel nitrate hexahydrate $Ni(NO_3)_2.6H2O$ are used as precursors of iron and other metals, whereas citric acid ($C_6H_8O_7$) is used as a complexant/fuel agent for the auto-combustion process. The required masses of the raw materials required to prepare the ferrite are shown in Table 1. These values are obtained using the following equation:

$$Wt\ (g) = M_w\ (g/mol) \times M\ (mol/L) \times V\ (L) \quad \ldots\ldots\ldots\ldots\ldots (1)$$

Where, Wt is the mass of the raw material, $M_w$ is the molecular weight of the raw material, M is the number of moles required for the material in one liter of solvent, and V is the volume of solvent.

Metal nitrates were entirely dissolved in small quantities of distilled water after being weighed. This solution was then mixed with citric acid to achieve a molar ratio of these nitrates and citric acid of 1:1 in the final sample. After that, ammonia is added to the mixture in droplets to balance the (pH) to (~7) while mixing it. Combustion reaction occurs among nearby metal nitrates and citrate molecules, resulting in a polymer network with colloidal dimensions recognized as sol [17-19]. While continuously mixing and heating the solution for one hour at 90 °C, the solution is evaporated, and then it held at this temperature until it solidified in a gel form. The gel then is cooked to 120 °C in order to trigger auto-combustion where the dried gel is burnt until it is totally consumed to produce loose powder. Finally, to get the required ferrite, the resultant powder is crushed in an agate mortar. The freshly as-prepared ferrite powder is then heated for two hours at 600 °C.

Table 1. The masses of raw materials required to obtain $Ni_xMn_{0.25-x}Mg_{0.75}Fe_2O_4$ ferrite.

| x | Composition | Ferric nitrate (g) | Magnesium nitrate (g) | Manganese nitrate (g) | Nickel nitrate (g) | Citric acid (g) |
|---|---|---|---|---|---|---|
| 0.00 | $Mn_{0.25}Mg_{0.75}Fe_2O_4$ | 32.32 | 7.6923 | 1.8900 | 0.00 | 23.0556 |
| 0.05 | $Ni_{0.05}Mn_{0.20}Mg_{0.75}Fe_2O_4$ | 32.32 | 7.6923 | 1.5120 | 0.5816 | 23.0556 |
| 0.10 | $Ni_{0.10}Mn_{0.15}Mg_{0.75}Fe_2O_4$ | 32.32 | 7.6923 | 1.1340 | 1.1632 | 23.0556 |
| 0.15 | $Ni_{0.15}Mn_{0.10}Mg_{0.75}Fe_2O_4$ | 32.32 | 7.6923 | 0.7560 | 1.7448 | 23.0556 |
| 0.20 | $Ni_{0.20}Mn_{0.05}Mg_{0.75}Fe_2O_4$ | 32.32 | 7.6923 | 0.3780 | 2.3264 | 23.0556 |

### 2.2. Fabrication of gas sensors

For each sample, 1.75 g of powder is collected and a pressure of 200 bar is applied by manual press for 120 seconds to produce a disc with a diameter of 1 cm and a thickness of 3.5 mm. The disc is then placed in furnace at a temperature of 900 °C for a period of two hours. Thin copper wires are used as connecting leads, and silver paste is used to construct the electrodes on one side of the sample, while electrodes are placed on all specimen surfaces to obtain Ohmic contacts [20]. The electrodes are fabricated for the five nano-ferrite samples, then the sensitivity of each sample to $NO_2$ gas at a constant concentration (65 ppm) is tested by a gas sensitivity test system.

### 2.3. Characterization

By using powder X-ray diffractometer (Philips PW1730), the ferrites' XRD (X-ray diffraction) pattern is obtained via Cu-Kα (Wavelength-1.5406 Å) radiation, scan range: 20º – 80º, and scan speed: 6 deg./min. The ferrites' surface morphology was investigated utilizing (MTRA3 LMU) field emission scanning electron microscope (FE-SEM) combined with Energy Dispersive X-ray Analyzer (EDX). A vibrating sample magnetometer (EZ VSM model 10) was used to measure the magnetism of some specimens. In order to detect ($NO_2$) gas at various temperatures, the gas response characteristics of sintered discs (900°C) were investigated. The resistance of gas sensor samples is measured by using Impedance Analyzer (UNI-TUT81B) equipped with a computerized testing tool.

## 3. Results and Discussion

### 3.1. X-Ray Diffraction

X-ray diffraction (XDR) analysis was carried out to determine the phase formation of the $Ni_xMn_{0.25-x}Mg_{0.75}Fe_2O_4$ nano-ferrite in the $2\theta$ range $10° \leq 2\theta \leq 80°$. Figure 1 shows the indexed x-ray diffraction patterns of the $Ni_xMn_{0.25-x}Mg_{0.75}Fe_2O_4$ ferrite annealed at 600 °C. The presence of (220), (311), (400), (422), (511), (440), and (533) planes confirms the formation of cubic spinel structure. The diffraction peaks agree with the JCPDS card number 89-3084 [21]. Additionally, the size of the crystallites gradually decreased as the amount of Ni doping increased. This was shown in the XRD pattern, where the $Ni_xMn_{0.25-x}Mg_{0.75}Fe_2O_4$ nano-ferrite peaks get shifted to higher angles, as the angle value increased, as listed in Table 2.

By using the Scherrer's equation, the crystallite size D of the $Ni_xMn_{0.25-x}Mg_{0.75}Fe_2O_4$ specimens was determined from the broadening of the (311) peak in the XRD patterns.

$$D = \frac{K\lambda}{\beta \cos\theta} \qquad \ldots\ldots\ldots\ldots\ldots\ldots (2)$$

Where, K is constant assumed to be 0.9, $\lambda$ is X-ray wavelength equal to 1.5406 (Å), $\beta$ is the full width at half maximum (FWHM) of the highest intensity diffraction peak expressed in radians, while $\theta$ is the Bragg's angle of the diffraction peak [22,23].

By using the following equation, the cubic unit cell lattice parameter (a) for all compounds was computed via diffraction planes:

$$a = d_{hkl}\sqrt{h^2 + k^2 + l^2} \qquad \ldots\ldots\ldots\ldots\ldots\ldots (3)$$

Where, d is the interplanar spacing and (*h, l and k*) are the Miller indices of the crystal planes [24]. The X-ray density ($\rho_x$) can be computed via the following equation:

$$\rho_x = \frac{8 M_W}{N_A a^3} \qquad \ldots\ldots\ldots\ldots\ldots\ldots (4)$$

Where, $M_W$ represents the molecular weight and $N_A$ is the Avogadro's number [25].

The lattice parameter (a), XRD density ($\rho_x$), and crystallite size (D) for all samples are given in Table 3.

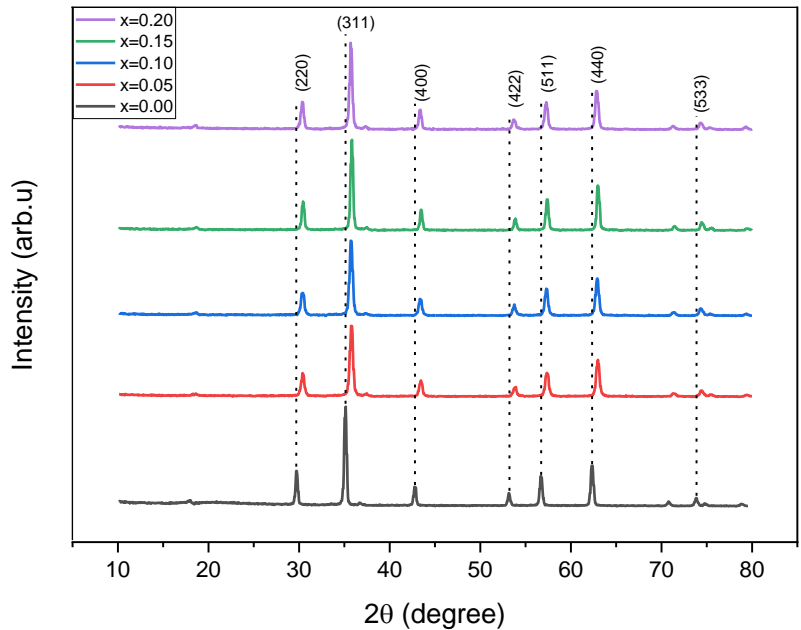

Figure 1. X-ray diffraction patterns of $Ni_xMn_{0.25-x}Mg_{0.75}Fe_2O_4$ nano-ferrite prepared by auto-combustion method.

Increasing the concentration of $Ni^{2+}$ leads to increase the lattice constant of ferrite compounds as listed in Table 3. Smaller $Fe^{3+}$ ions have been observed to migrate from tetrahedral to octahedral positions in response to $Ni^{2+}$ addition [26,27], therefore tetrahedral sites are enlarged as a result of increasing the lattice constant [28,29]. Moreover, this caused the lattice to grow and the density to drop, indicating that the lattice constant has changed as a result of the dopant ions being absorbed into the lattice could have taken an interstitial positions among the hosting ions [20].

Table 2. Structure properties of the $Ni_xMn_{0.25-x}Mg_{0.75}Fe_2O_4$ nano-ferrite.

| *h k l* | 2θ (deg) (JCPDS) | 2θ (deg) (x=0.00) | 2θ (deg) (x=0.05) | 2θ (deg) (x=0.10) | 2θ (deg) (x=0.15) | 2θ (deg) (x=0.20) |
|---|---|---|---|---|---|---|
| **220** | 30.115 | 30.1365 | 30.4563 | 30.3111 | 30.3932 | 30.3938 |
| **311** | 35.466 | 35.4950 | 35.8238 | 35.7308 | 35.8876 | 35.7541 |
| **400** | 43.123 | 43.2299 | 43.5441 | 43.4461 | 43.4725 | 43.3345 |
| **422** | 53.478 | 53.5835 | 53.9189 | 53.7877 | 53.8403 | 53.6563 |
| **511** | 57.000 | 57.1528 | 57.4708 | 57.3573 | 57.4057 | 57.2337 |
| **440** | 62.594 | 62.7239 | 62.8946 | 62.9067 | 62.9564 | 62.8185 |
| **533** | 74.049 | 74.2529 | 74.3735 | 74.2861 | 74.3755 | 74.2936 |

Table 3. Unit cell constant (a), density ($\rho_x$) and crystallite size (D) of $Ni_xMn_{0.25-x}Mg_{0.75}Fe_2O_4$ nano-ferrite prepared by auto-combustion method.

| x | Composition | a (Å) | $\rho_x$ (g/cm³) | D (nm) |
|---|---|---|---|---|
| **0.00** | $Mn_{0.25}Mg_{0.75}Fe_2O_4$ | 8.36743 | 5.250 | 28.31 |
| **0.05** | $Ni_{0.05}Mn_{0.20}Mg_{0.75}Fe_2O_4$ | 8.37691 | 5.232 | 24.34 |
| **0.10** | $Ni_{0.10}Mn_{0.15}Mg_{0.75}Fe_2O_4$ | 8.38131 | 5.224 | 24.34 |
| **0.15** | $Ni_{0.15}Mn_{0.10}Mg_{0.75}Fe_2O_4$ | 8.38245 | 5.222 | 28.32 |
| **0.20** | $Ni_{0.20}Mn_{0.05}Mg_{0.75}Fe_2O_4$ | 8.38717 | 5.213 | 24.30 |

### 3.2. FE-SEM and EDX Analysis

To assess the morphology of the fabricated samples, (FE-SEM) was used. Figure 2 illustrates the $Ni_xMn_{0.25-x}Mg_{0.75}Fe_2O_4$ nano-ferrite micro images at a 200 nm scale after annealing at 600 °C. The observed FE-SEM images made it extremely apparent that the magnetic ferrite particles were created through some aggregation at the nanoscale. The FE-SEM images show porous, sponge-like shape particles of the samples (x = 0.00, and 0.05). Most likely, the gases released during the gel's combustion process are what caused the pores to form [30]. In addition, the images show particles that are spherical or semi-spherical and nonhomogeneous in form of the samples (x=0.10, and 0.15), as well as the images show homogeneous distribution and spherical nanoparticles of the sample (x = 0.20). The FE-SEM images also show the formation of tiny agglomerated grains with surface spaces or voids and no distinct shape. The agglomerates are where the porosity is located. Since gas detecting is a surface phenomenon and porosity is essential, the reported porous microstructure is beneficial for sensing purposes [31]. It is obviously shown in the micrographs that the particles structures of the $Ni_xMn_{0.25-x}Mg_{0.75}Fe_2O_4$ nano-ferrite are very coarse, which facilitate adsorption of oxygen species on the detecting surface. Adsorption of oxygen species is responsible for gas detecting [32].

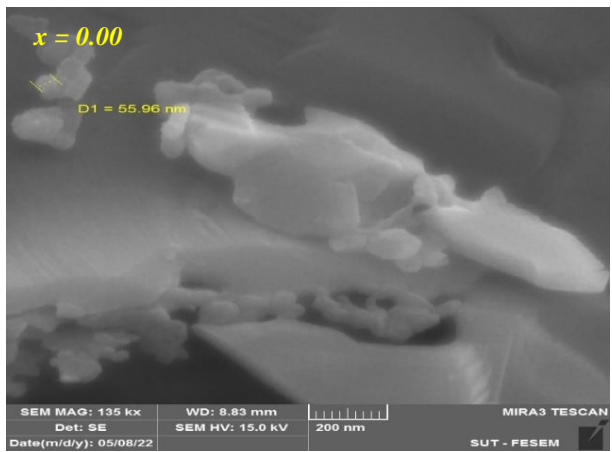 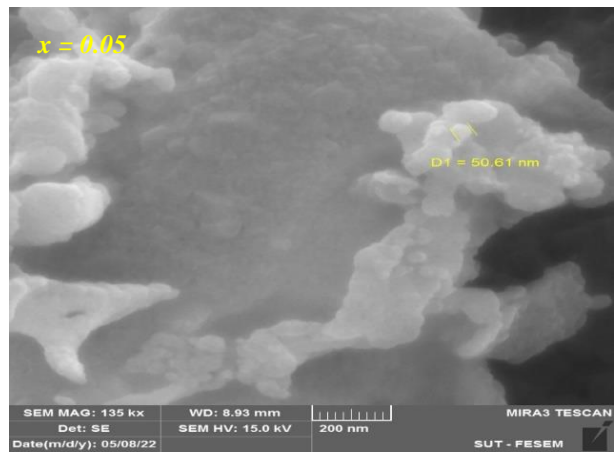

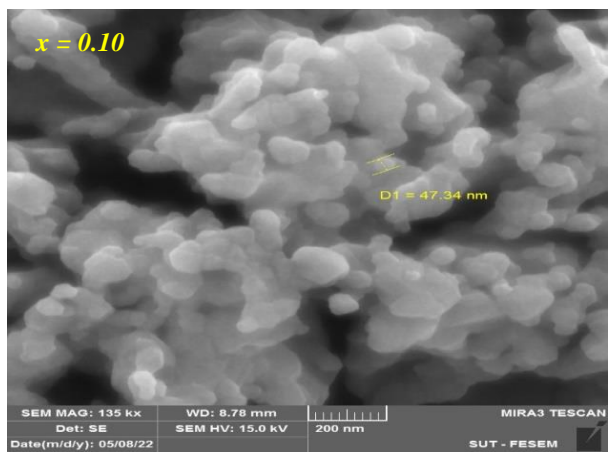 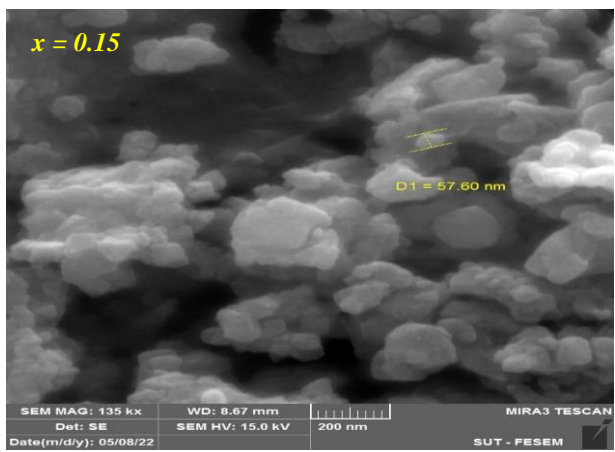

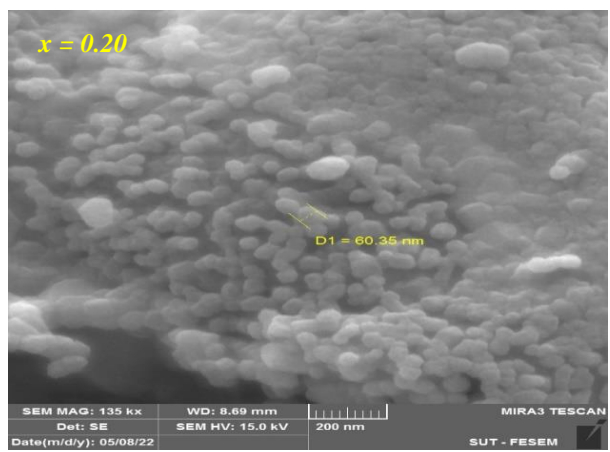

Figure 2. FE-SEM images of $Ni_xMn_{0.25-x}Mg_{0.75}Fe_2O_4$ nano-ferrite.

The EDX spectra of the $Ni_xMn_{0.25-x}Mg_{0.75}Fe_2O_4$ nano-ferrite (where x = 0.00, 0.05, 0.10, 0.15 and 0.20) are illustrated in Figure 3, referring that the spectral lines related to (Ni, Mn, Mg, Fe and O), verify that the synthesized compound $Ni_xMn_{0.25-x}Mg_{0.75}Fe_2O_4$ was achieved.

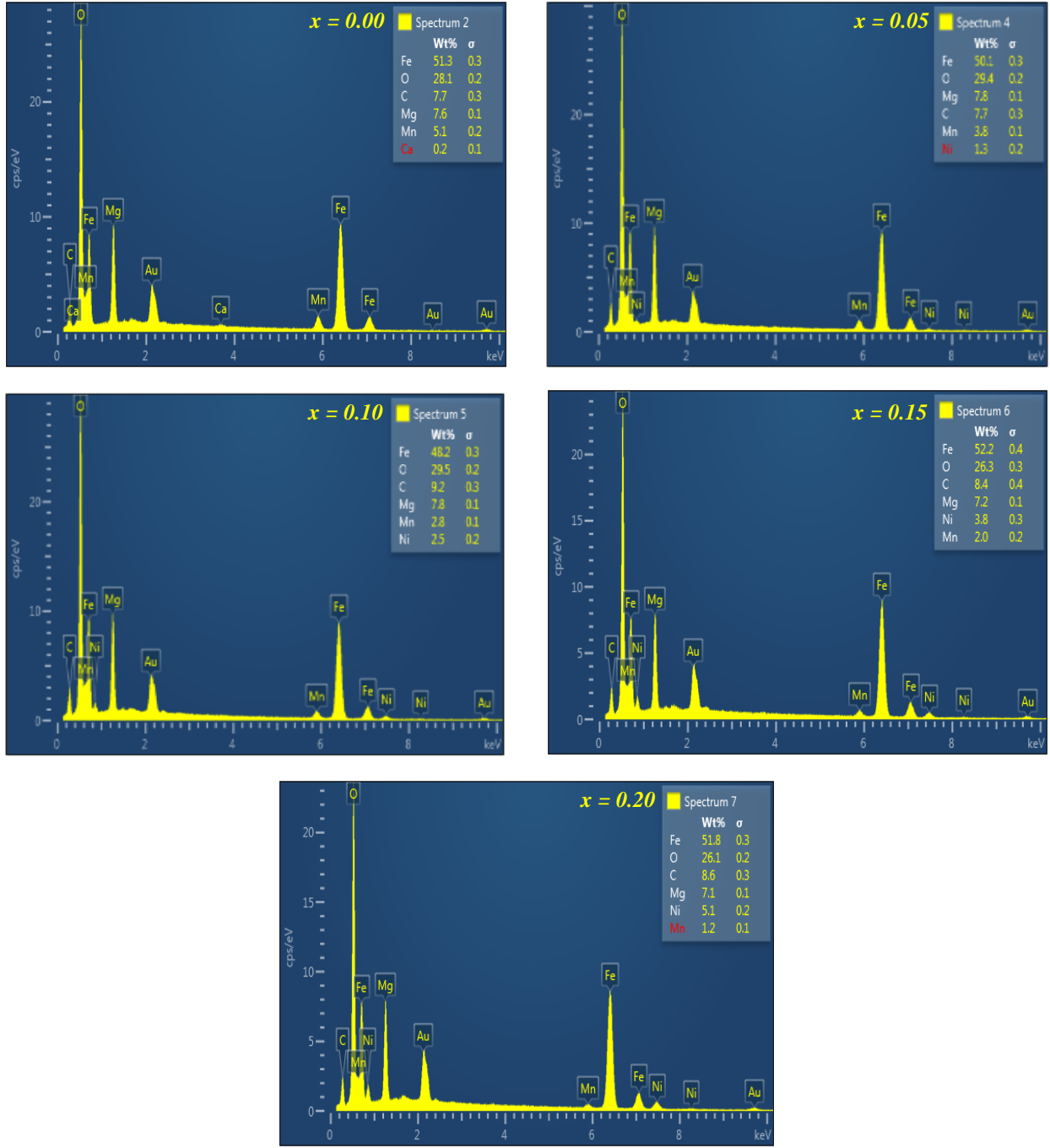

Figure 3. EDX spectra of $Ni_xMn_{0.25-x}Mg_{0.75}Fe_2O_4$ nano-ferrite.

## 3.4. Magnetic Characteristics

Hysteresis loop is measured utilizing a (VSM) system, and magnetic characteristics of samples were examined at room temperature (300 K). Figure 4 shows the hysteresis loop curves of $Ni_xMn_{0.25-x}Mg_{0.75}Fe_2O_4$ (x = 0.00, and 0.20). (S) shaped curves indicate that standard soft magnetic material and magnetic coercivity can be ignored. In addition, the particles are so small that they behave like superparamagnetic material. Due to the small crystallite size, as is evidenced by the XRD analysis in Table 3, nanoparticles have superparamagnetic behavior, in which their magnetic moments attempt to align with one another in a specific way [33,34].

According to Neel, the distribution of cations among the octahedral and tetrahedral locations in spinel ferrite determines the overall magnetic moment [35]. Saturation magnetization ($M_s$), remnant magnetization ($M_r$), and magnetic coercivity ($H_c$) values were computed from the M-H curves depending on ($M_s$) measured values.

M-H curves have demonstrated how chemical compound affects magnetic properties. Table 4 illustrates the variation in saturation magnetization ($M_s$) values for specimens captured from hysteresis loop curves. As 0.20 of the $Ni^{2+}$ ions were swapped out for $Mn^{2+}$ ions, the $M_s$ value dropped from 28.980 (emu/g) for x = 0.00 to 23.400 (emu/g). According to experimental observations, as nickel content rises, the ratio of ferric, manganese, or magnesium ions on the A-location decreases, while at the same time, the of $Fe^{3+}$ ions grows by the same amount on the location B. As a result, the A-B interaction is reduced. As a consequence of the ionic moments on the B-sites no longer being maintained parallel to each other, the angles among them start to form, which lowers the moment of the B sub lattice itself. Most likely, nickel ions have been replaced by cations in the B-sites [34]. Figure 4 shows how the observed values of the remnant magnetization ($M_r$) and coercive field ($H_c$) are so small, demonstrating that the grain size does not pass the critical diameter of single-domain grain [34]. The cation distribution has a significant impact on the net magnetic moments and magnetocrystalline anisotropy. Table 4 lists the magnetic factors.

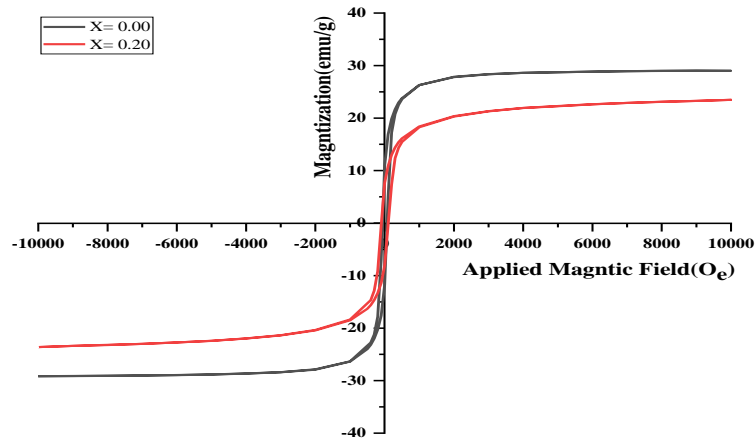

Figure 4. Magnetization (M) versus applied magnetic field ($O_e$) of $Ni_xMn_{0.25-x}Mg_{0.75}Fe_2O_4$ (x = 0.00, and 0.20) nanoparticles at 300K.

Table 4. Variation of magnetic factors for $Ni_xMn_{0.25-x}Mg_{0.75}Fe_2O_4$ (x =0.00, and 0.20) nanoparticles.

| x | Compound | $M_s$ (emu/g) | $M_r$ (eum/g) | $H_c$ ($O_e$) |
|---|---|---|---|---|
| **0.00** | $Mn_{0.25}Mg_{0.75}Fe_2O_4$ | 28.98 | 10.95 | 61.50 |
| **0.20** | $Ni_{0.20}Mn_{0.05}Mg_{0.75}Fe_2O_4$ | 23.40 | 7.54 | 94.00 |

### 3.3. Gas Sensing Features

The gas concentration, material composition, type of conductivity, operating temperature, and different controlling parameters are considered as important factors which affect the gas sensitivity or gas response of the metal oxide semiconductor sensor [36]. Depending on the compound and operating temperature, the gas sensitivity of the $Ni_xMn_{0.25-x}Mg_{0.75}Fe_2O_4$ (where x= 0.00, 0.05, 0.10, 0.15, and 0.20) nano-ferrite against $NO_2$ gas is studied and computed using following equation:

$$S = \left|\frac{Rg-Ra}{Ra}\right| \times 100\,\% \qquad \text{[Oxidizing gas]} \qquad \ldots\ldots\ldots\ldots\ldots\ldots (5)$$

Where $R_g$ and $R_a$ represent the electrical resistances in the $NO_2$ gas and air, respectively [37, 38].

Figure 5 shows the sensing characteristics and variation for each sample against nitrogen dioxide $NO_2$ gas when exposed and removed the examined gasses of the $Ni_xMn_{0.25-x}Mg_{0.75}Fe_2O_4$ nano-ferrite. As can be seen from the figure, the resistance value increases when the discs are exposed to $NO_2$ gas (Gas ON), and subsequently decreases when the gas is closed (Gas OFF) for all samples. At concentration of 65 ppm of $NO_2$, the sensor's sensitivity was examined at various operating temperatures (200 °C, 250 °C, and 300 °C). In the existence of an oxidizing gas, the operating temperature is required to change the material's oxidation state and the conductivity of $Ni_xMn_{0.25-x}Mg_{0.75}Fe_2O_4$ nano-ferrite. The response time is defined as the amount of time needed to reach 90% of the equilibrium response of the gas, while the recovery time, is defined as the amount of time needed to reach 10% of the baseline resistance [39]. From Table 5, it can be seen that samples demonstrate a high sensitivity to nitrogen dioxide gas at 250 °C while it is around 300 °C for sample x=0.00. As shown in the FE-SEM images, the sensitivity of the doped samples increases because it has the highest roughness, and this is agreement with the findings of researchers [20,32]. Additionally, the figure also demonstrates that the $Ni_{0.20}Mn_{0.05}Mg_{0.75}Fe_2O_4$ ferrite compound has its highest gas response 707.22% of the sample (x=0.20) at 250 °C. Since the sensitivity process in metal oxides occurs through the adsorption of oxygen ions on the surface, doping of Mn by Ni generally often enhances the sensitivity because a lack of oxygen causes the formation of oxygen voids; (When the oxygen concentration in the $Ni_xMn_{0.25-x}Mg_{0.75}Fe_2O_4$ lattice increases, more oxygen ions ($O^{-2}$ and $^-O$) adsorb to the sensor's surface due of the gaps or voids) [20]. In contrast to the pre-adsorbed oxygen and other test gases, $NO_2$ gas has a greater electron affinity and is a very reactive and oxidizing gas [40]. After the covalent bond between nitrogen and oxygen is formed, $NO_2$ has an unpaired electron, and remains as one of the atoms with a single unpaired electron. Because the nano-ferrite has a short response time (1.2-11.4) s at 200 °C and a short recovery time (1.5-4.4) s at 250 °C, it is possible to conclude that the sensor has excellent sensing characteristics. This fast response of the sensor could be a result of the small particle size, which causes the particle boundaries to enlarge. The values of sensitivity, response time, and recovery time are tabulated in Table 5.

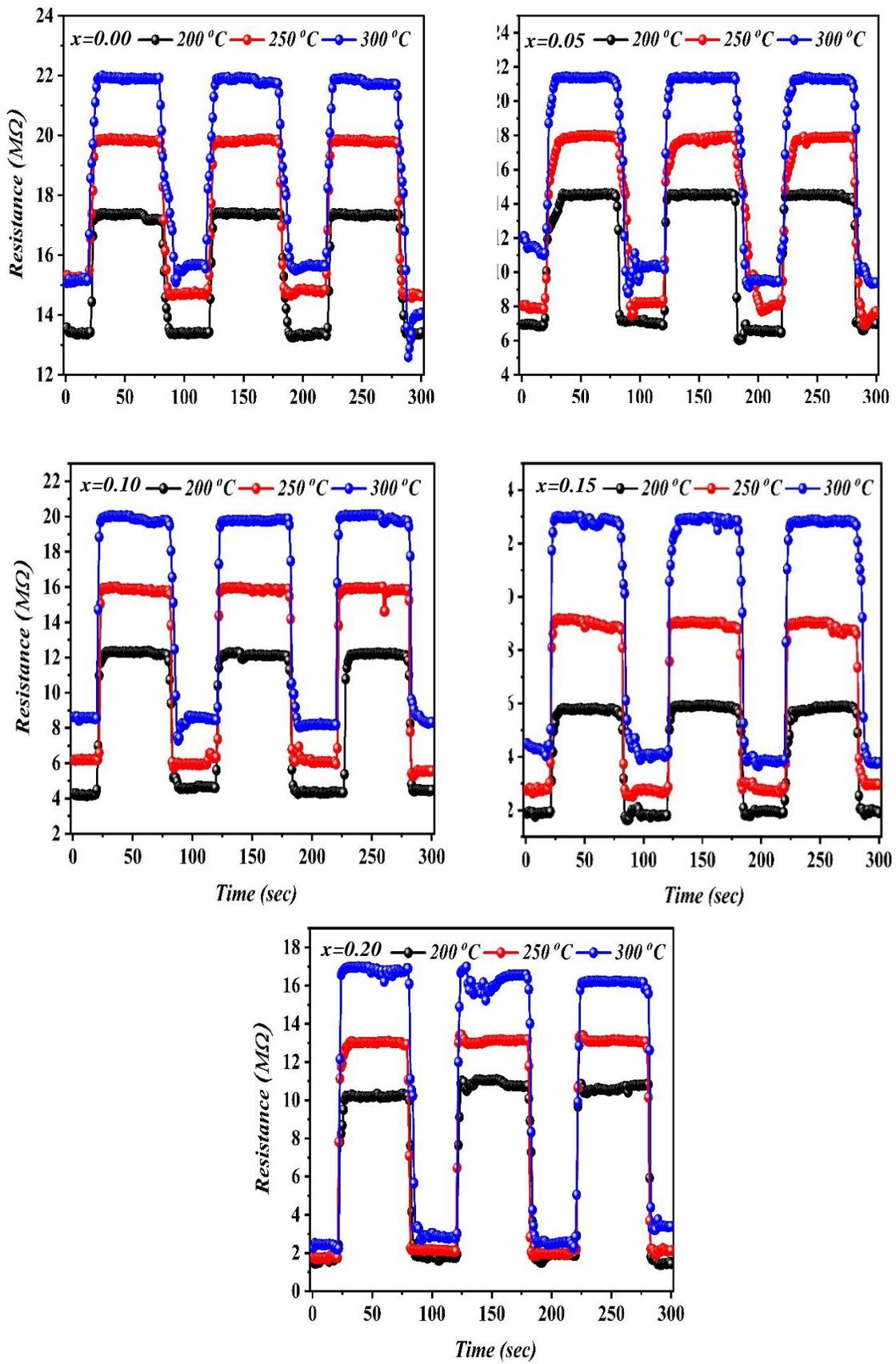

Figure 5. The variation in resistance with time of Ni$_x$Mn$_{0.25-x}$Mg$_{0.75}$Fe$_2$O$_4$ nano-ferrite at different operating temperatures.

Table 5. NO$_2$ gas sensitivity, response time and recovery time values of Ni$_x$Mn$_{0.25-x}$Mg$_{0.75}$Fe$_2$O$_4$ nano-ferrite at different operating temperatures.

| x | Response Time | | | Recovery Time | | | Sensitivity % | | |
|---|---|---|---|---|---|---|---|---|---|
| | 200 °C | 250 °C | 300 °C | 200 °C | 250 °C | 300 °C | 200 °C | 250 °C | 300 °C |
| **0.00** | 2.4 | 4.0 | 5.9 | 5.2 | 4.4 | 11.0 | 30.82 | 36.30 | 74.60 |
| **0.05** | 11.4 | 11.4 | 5.5 | 1.9 | 1.9 | 6.3 | 141.72 | 160.11 | 134.45 |
| **0.10** | 2.0 | 1.5 | 1.9 | 3.6 | 1.5 | 4.7 | 198.07 | 202.45 | 175.34 |
| **0.15** | 11.4 | 3.2 | 9.0 | 9.7 | 3.0 | 9.6 | 262.80 | 264.28 | 255.22 |
| **0.20** | 1.2 | 3.7 | 1.63 | 1.8 | 2.3 | 5.24 | 707.34 | 707.22 | 676.25 |

## 4. Conclusions

Utilizing a simple sol-gel auto-combustion process, Ni$_x$Mn$_{0.25-x}$Mg$_{0.75}$Fe$_2$O$_4$ nano-ferrite was synthesized using metal nitrates as a source of cations and citric acid (C6H8O7) as a complexant/fuel agent for the auto-combustion process. The Ni$_x$Mn$_{0.25-x}$Mg$_{0.75}$Fe$_2$O$_4$ nano-ferrite with the spinel structure peaks in the XRD patterns corresponding to the investigated systems, and no unidentified peaks are observed. The FE-SEM images show microstructures with open pores and nanoscale grains with agglomeration, which is nearly comparable to the crystalline size determined by XRD. These findings reveal that, due to the particles being small, the prepared samples at-room-temperature hysteresis loop curves exhibit superparamagnetic behavior. Furthermore, the results of the NO$_2$ gas sensing showed that the gas sensor had a good performance in terms of its response to the gas. The sensitivity increases with the increasing concentration of Ni in composition, as well as it also boasts shorter response and recovery times. For gas sensing applications, in Mn$_{0.25}$Mg$_{0.75}$Fe$_2$O$_4$ it is concluded that it is desirable to substitute manganese ions by nickel ions.